# Beam Phase Retrieval based on Higher Order Modes in Cylindrical SRF Cavities

L. Shi, [1,2,3,a)] N. Baboi,[2] and R.M. Jones[1,3]

[1]*School of Physics and Astronomy, the University of Manchester, Manchester, M13 9PL, UK*

[2]*Deutsches Elektronen-Synchrotron (DESY), Notkestrasse 85, 22607 Hamburg, Germany*

[3]*The Cockcroft Institute of Accelerator Science and Technology, WA4 4AD, Daresbury, UK*

The control of beam phase relative to the accelerating RF field within a superconducting cavity is important in many accelerator applications and is of particular importance for a free electron laser facility. As standard practice, the phase is usually inferred from the beam-induced transient field with respect to a timing reference. We report here on an alternative and novel means of beam phase determination based on beam-excited higher order electromagnetic modes and the accelerating electromagnetic mode, which are conveniently available from the same coupler. The monopole modes are immune to the electron beam offset and therefore are best suited for the task. A coupled circuit model is used to assist the development and to rapidly assess the facility of the method. Simulations based on the circuit model indicate that the resolution of this system depends critically on the signal to noise ratio. Beam-based measurements with a test setup were carried out at the European XFEL, Germany. Based on this new method we have routinely obtained a resolution of 0.1°. The best resolution observed with the current setup was 0.03°. These results agree very well with the predictions from those predicted by a circuit model. The system investigated here can be used to provide diagnostic information for the current LLRF system employed in the European XFEL. To this end, the associated electronics are under development. This monitor is the first of its kind that can deliver direct and online measurements of the beam phase with respect to the RF field.

## I. INTRODUCTION

The E-XFEL (European X-ray Free Electron Laser) [1] is hosted in Hamburg and Schenefeld. It is a SASE-FEL (Self Amplified Spontaneous Emission Free Electron Laser) facility that aims to produce high quality hard X-rays, which achieves laser amplification and saturation with a single passage of electron bunches through hundreds of meters long undulator sections. The electron bunches have a repetition rate of 4.5 MHz within each RF pulse. The RF pulses are operated with a 10 Hz rate. At the end of the 1.7 km long linac, the electron energy reaches up to 17.5 GeV. The electron beam traverses undulators and the motion of the beam within this region generates photons with wavelength down to ~0.1 nm range. The wavelength in the X-ray range has the potential to explore physics processes at atomic scales while the high photon intensity can be used to create extreme conditions of high pressure, temperatures or electromagnetic fields [1].

Along the 3.4 km long facility, the majority of the space is occupied by 97 accelerating cryomodules. Each module contains eight superconducting SRF cavities. These nine cell cavities are known as TESLA (TeV Energy Superconducting Linear Accelerator) cavities (see Fig. 1) [2]. Four cryomodules are regulated by one RF power station. A digital LLRF (Low Level RF)

a) Electronic mail: liangliang.shi@psi.ch, now at Paul Scherrer Institut

system is used to regulate the RF amplitude and phase. The amplitude and phase stability requirements for the European XFEL are 0.01% and 0.01° respectively [3].

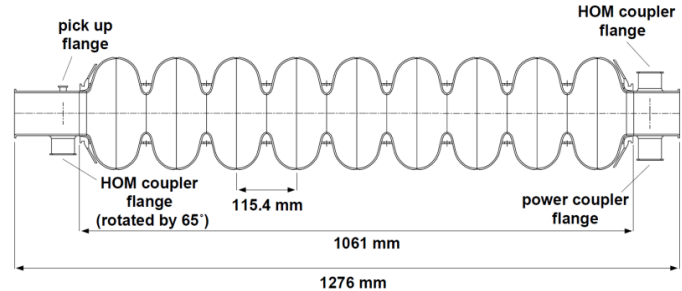

**FIG. 1. TESLA cavity with dimensions. There are three couplers: the power coupler for RF power input, two HOM couplers (the left HOM coupler is rotated by 65° with respect to the right one for displaying) for HOM damping as mentioned in section II. In addition there is also a RF pickup for RF diagnostic [2].**

Our study focuses on the injector part at the E-XFEL, because the experience from FLASH [4] shows that the beam phase at the injector part is critical for the FEL beam quality [5].

**II. Wakefields and Higher Order Modes**

When a charged particle, or indeed a bunch of electrons traverses a cavity, it self-excites an electromagnetic field which can be conveniently represented in terms of a wakefield. This field can seriously degrade the electron beam quality if left unchecked [6] and, in the worst case, can cause BBU (Beam Breakup) [7]. The wakefield can be decomposed in a multipole expansion of modes. In general, provided the beam is not appreciably offset from the electrical center, the monopole and dipole modes are dominant [7]. The modes with higher resonant frequencies than the accelerating mode are referred to as HOMs (Higher Order Modes). These modes are grouped into bands for multi-cell structures. In order to avoid their damaging effects, these HOMs have to be suppressed through attenuating them directly or detuning them [6]. The TESLA cavities are equipped with HOM couplers which attenuates the wakefield. These damped HOMs carry useful information about the electron beam and therefore can be used for beam diagnostics. For this purpose we sample a small portion of these modes.

For a cavity that exhibits cylindrical symmetry around its axis, the beam-excited monopole modes are independent of the beam offset and are only proportional to the beam charge. It is this characteristic that ensures that the signal strength of the monopole modes is immune to the beam orbit variations, which is attractive for the application for beam phase monitoring.

We note that in practice the symmetry is broken by HOM, fundamental power couplers, inevitable manufacturing errors etc. For TESLA cavities the perturbation from these effects is small [19] so that the monopole mode characteristics are maintained. The first monopole band contains the 1.3 GHz mode used for acceleration, and we utilize the second monopole band (~2.4 GHz) for monitoring the phase of the beam [8]. We also use the first dipole band to serve as a beam position monitor, but this is not the focus of this paper as we have reported it elsewhere [9][10].

**III. HOM-based Beam Phase Determination**

The use of HOMs for beam phase measurements was mentioned in [9]. Each beam induced mode carries the arrival time information of the beam. By measuring the signal available at the HOM coupler, it is possible to convert the timing information into beam phase relative to the 1.3 GHz RF field inside the cavity measured from the same port. The modes appropriate for this task are in the first monopole HOM band at around 2.4 GHz. The procedure is described in detail in Appendix A.

This paper is structured such that, in section A we develop a coupled circuit model and this is employed to guide the beam phase monitor development. In section B, the experimental measurement, based on a broadband setup, made at the E-XFEL injector is reported. The measurement and simulation results are presented in section C together with some conclusions on the method.

### A. Coupled Circuit Model

A coupled resonant circuit model (based on the one originally developed by D. Nagle, E. Knapp and B. Knapp [11]) is used to study the dynamics of the second monopole band. The circuit model is schematically shown in Fig. 2. Each parallel resonance circuit unit describes one cell of the cavity and contains one capacitor $C$, two inductors $2L$, and a current source $I$. The capacitor and the inductor can be identified with the RF cavity parameters as described in [12]. In order to study the beam progress in the cavity, each unit of the circuit is driven by a Gaussian pulse. The time delay of two adjacent pulses is set to half period of 1.3 GHz mode, which is ~0.38 ns. The normalized voltage across each capacitor can then be obtained by solving the circuit, as described in [12]. The circuit model is implemented in Simulink® and solved with a 5 pico-second step. The voltages across the first and ninth units, denoted by HOM1 and HOM2 respectively, are superimposed with one at the accelerating frequency $f_{acc}$ (=1.3 GHz) to simulate the signals available from HOM ports.

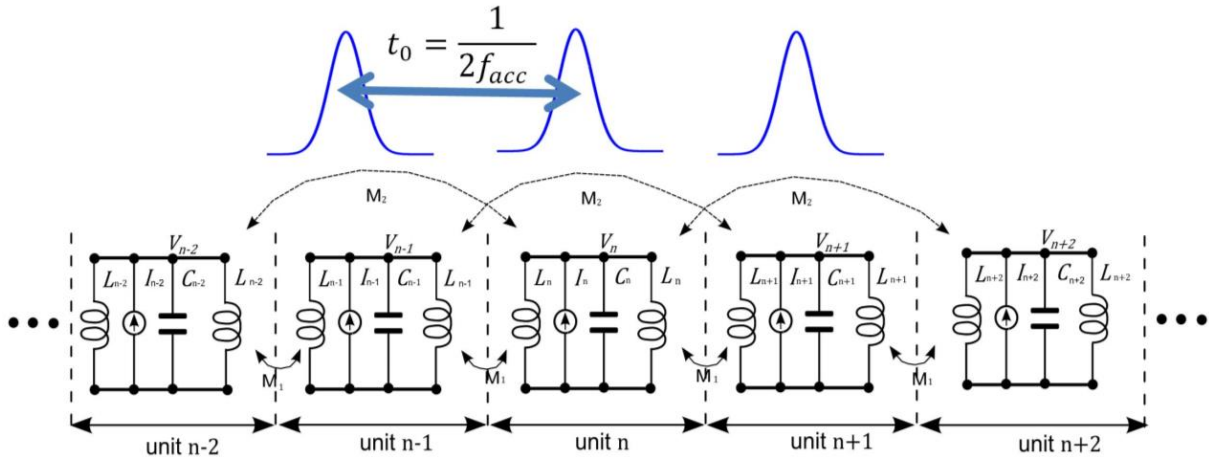

**FIG. 2. A coupled circuit model driven by propagating Gaussian pulses.**

We drive the circuit with these Gaussian pulses and record the voltages across each parallel chain. In Fig. 3, an example spectra corresponding to HOM1 and HOM2 are shown. Also displayed are vertical lines which demarcate the eigenmode frequencies of the TESLA cavity and these have been obtained with simulations using the MAFIA code [13]. There are nine modes excited corresponding to the nine eigenmodes of the 2$^{nd}$ monopole band. These modes are denoted by mode 1, mode 2 … and mode 9 according to their frequencies in ascending order.

Mode 8 and mode 9 are excited stronger due to their higher R/Q relative to others [13]. Therefore, these two modes are used to determine the beam phase. In principle, nine modes can be used jointly to give the beam phase, but the improvement in terms of resolution is negligible, whilst the computation power required is tripled.

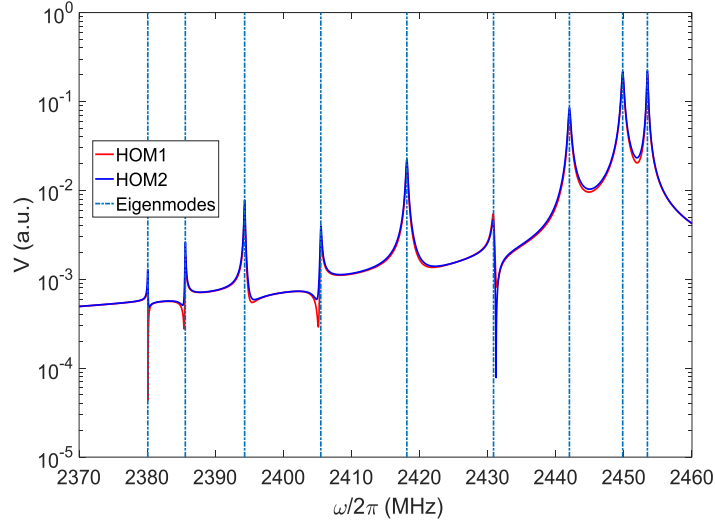

**FIG. 3. Spectra of voltage waveforms HOM1 and HOM2 in the vicinity of 2.4 GHz, as obtained from circuit model. The vertical dashed lines are the eigenmodes frequencies from a MAFIA simulation of the TESLA cavity [13].**

The phase of the 1.3 GHz signal can be varied to simulate a RF phase change and then can be retrieved based on the procedure described in Appendix A. The calculation is performed independently for HOM1 and HOM2. The RMS of the phase difference between HOM1 and HOM2 is used to evaluate the resolution of the beam phase determination. A factor of $\frac{1}{\sqrt{2}}$ is applied to estimate the resolution of each channel by assuming that the two channels have the same performance. A variable level of AWGN (Additive White Gaussian Noise) is superimposed to the HOM1 and HOM2 to simulate the experimental data. The phase of the 1.3 GHz signal is varied by ±5°. HOM1 and HOM2 are sampled at different sampling rates to investigate their resolution dependence. The results are summarized in Fig. 4. Also displayed is an exponential fit to the data of the form

$$\Delta\theta = ae^{-mx}, \tag{1}$$

where $\Delta\theta$ is the resolution (in degrees) and $a$, $m$ are parameters to be determined.

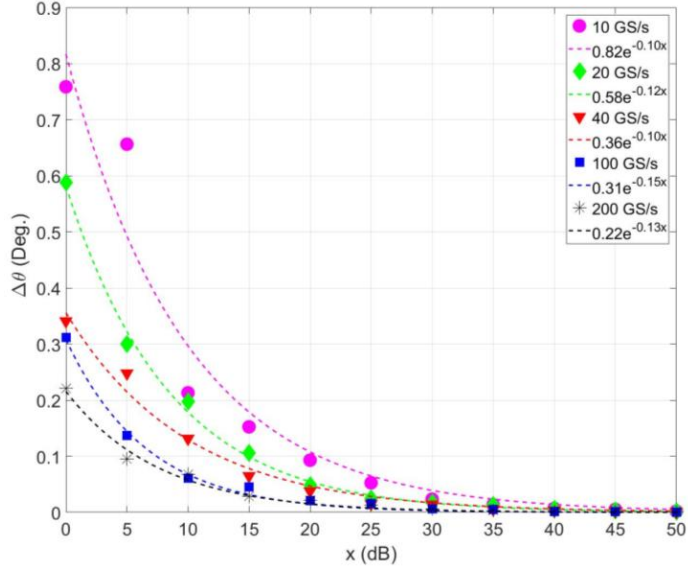

**FIG. 4. Resolution Δθ versus signal to noise ratio ($x$) for different sampling rates. The SNR is varied from 0 to 50 dB in a step of 5 dB. The data is sampled at 10, 20, 40, 100, and 200 GS/s. An exponential fit to each set of data is also shown.**

The parameters associated with these fits are summarized in Table I.

**Table I Summary of fitting parameters used in Fig. 4.**

| Sampling Rate (GS/s) | $a$(Deg.) | $m$ | $R^2$ |
|---|---|---|---|
| 10 | 0.82 | 0.10 | 0.947 |
| 20 | 0.58 | 0.12 | 0.997 |
| 40 | 0.36 | 0.10 | 0.986 |
| 100 | 0.31 | 0.15 | 0.995 |
| 200 | 0.22 | 0.13 | 0.988 |

Here $R^2$ is the coefficient of determination and is representative of how good the exponential fit is to the data.

A higher sampling frequency will provide an enhanced resolution but it is of course ultimately limited by the SNR. For a sampling rate of 20 GS/s, as used later in the measurements, $a$ and $m$ are 0.58 and 0.12 respectively. Given a SNR of 35 dB, we find equation 1 predicts a resolution Δθ of 0.009°, which meets the 0.01° phase requirement at the E-XFEL [14].

**B.    Experimental setup and measurements**

Beam phase measurements were made at the E-XFEL injector module using a fast oscilloscope. The experimental setup is shown in section B.1. Fig. 4 indicates that the resolution has a strong dependence on the SNR. Therefore, in order to compare the simulation to the experimental results, the SNR is estimated. We used SVD (Singular Value Decomposition) to separate the real signal and noise, which is presented in section B.2.

**B.1 Experimental setup**

The experimental setup consists of several RF bandpass filters, combiner/splitter, and a fast Tektronix® scope TDS6604B (20 GS/s with 6 GHz bandwidth). The setup is shown schematically in Fig. 5.

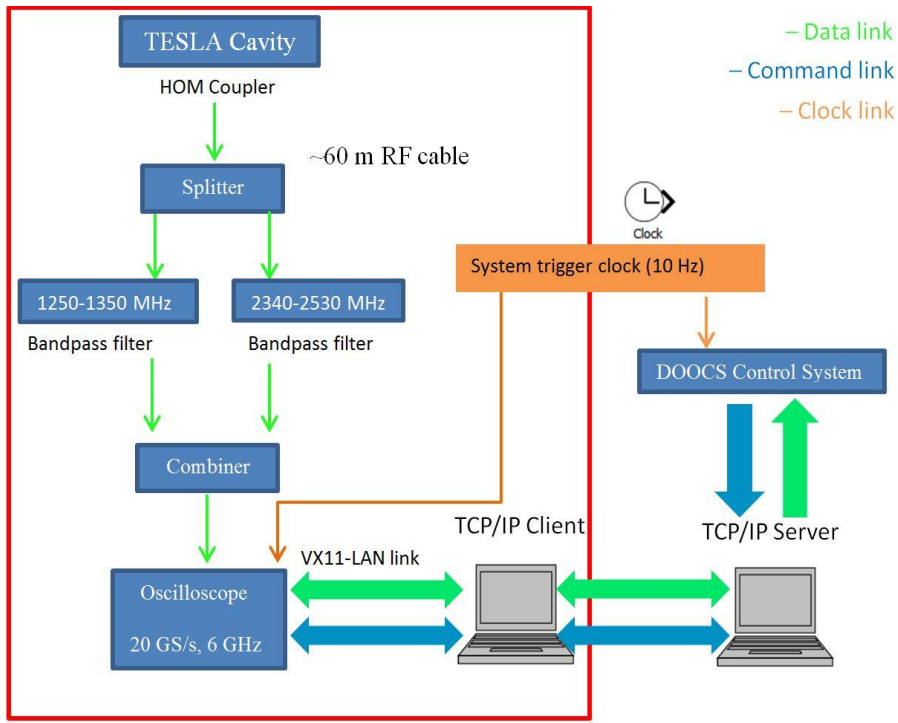

**FIG. 5. Block diagram of the beam phase measurement setup.**

The two HOM couplers of each cavity deliver the signals for the two channels used for beam phase measurements. The signal from each HOM coupler is transmitted by a RF cable with length of ca. 60 m from the tunnel to the measurement rack. The signal is then split with a Mini-Circuit® power splitter (5-2500 MHz). Each split signal is filtered, one centered at approximately 1300 MHz with 100 MHz bandwidth and the other approximately 2435 MHz with 190 MHz bandwidth. The filtered signals are then combined again before they are fed into the fast scope. The scope is triggered by an external 10 Hz trigger, which is synchronized with the data acquisition by the DOOCS [15] control system. The scope is remotely connected to a PC with VX11 protocol. One PC serves as a TCP/IP client and a second as a server for collecting data from DOOCS. It should be pointed out that the whole system is only partially synchronized with the electron beam because the synchronism is provided by the TCP/IP protocol and the command routing inside the network depends on the momentary traffic. It takes approximately 20 seconds to complete one triggered measurement.

As an example, the waveforms and the associated spectra from both HOM couplers are shown in Fig. 6. The waveform is 20 µs long and the frequency step in the spectrum is 50 kHz.

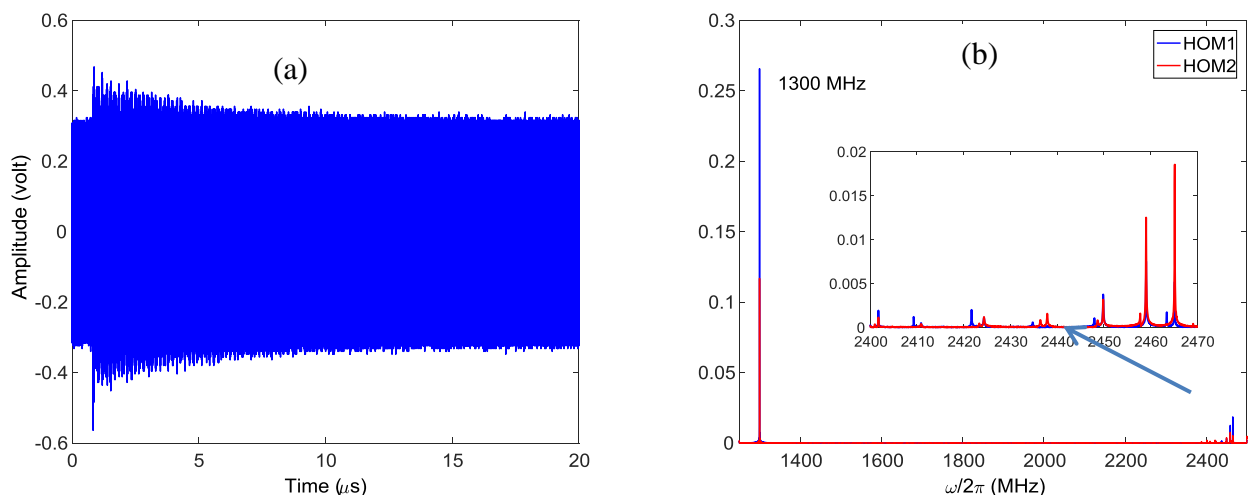

**FIG. 6. Measured waveforms (a) and spectra (b) from HOM 1 (blue) and 2 (red). The inset shows the spectra in the region of 2.4 GHz. The last two modes are excited strongly, and are used for the phase determination.**

The last two modes (mode 8 and mode 9) in the spectrum are excited strongly. The mode frequencies can be easily identified by the overlapping of the signals from HOM1 and HOM2.

### B.2 Non-parametric noise filtering and estimation

As shown in Fig. 4, the resolution has a strong dependence on the noise level. Therefore, it is important to estimate the noise level in the measurements. For this purpose, we used a SVD [20] (Singular Value Decomposition) method. SVD is a powerful model independent method to extract features from data or images and therefore reduce the dimension of the dataset. We apply it here for signal decomposition. By arranging the signals in a matrix row wise, we form a data matrix $D$. The SVD eigen components of the matrix $D$ can be found in $V$, which are the basis for the signal decomposition,

$$D = USV^T, \qquad (2)$$

where $U = [u_1, \dots, u_m]$ and $V^T = [v_1, \dots, v_n]$ is the transposed matrix of $V$. The singular matrix $S$ contains information about the signal amplitude and noise etc. A reconstructed signal matrix $D_k$ is then obtained based on the first k components,

$$D_k = \sum_{i=1}^{k} s_{ii} u_i v_i^T, k \leq \text{rank}(D). \qquad (3)$$

The signals show larger singular values due to the correlation among them, while the noise only shows smaller singular values. By using a suitable number $k$ in equation 3, the signal and the noise can be separated. The separation is not perfect based on this method, but it provides an estimation of the noise level in the measurements. The measurements were made at the second cavity in the injector module A1 at the E-XFEL. Seventy five measurements were used to form the data matrix $D$. The singular values of the data from HOM1 and HOM2 are shown in Fig. 7.

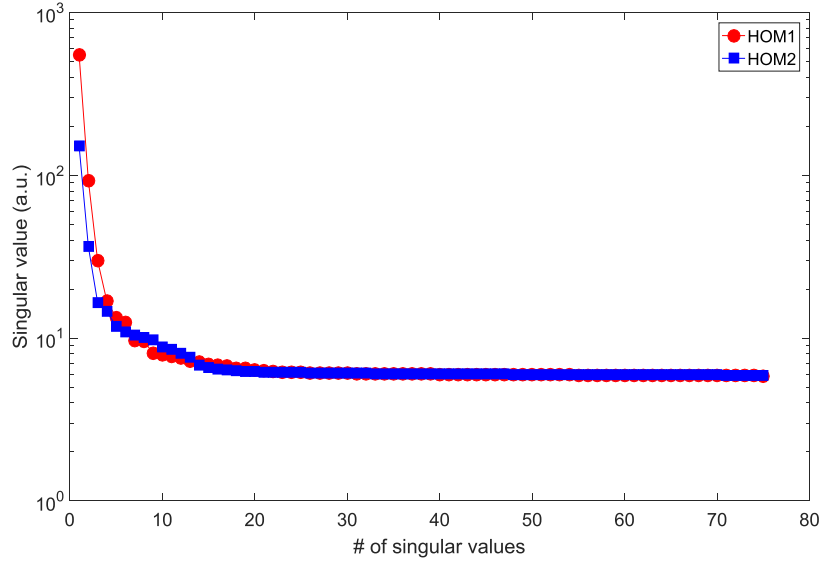

**FIG. 7. Singular value distributions for the *D* matrices for HOM1 and HOM2. The singular values from both channels drop quickly to the same level.**

The first 24 singular values were used in the reconstruction of the signal and the rest is regarded as noise. The noise waveform is reconstructed based on these remaining 51 singular values. The waveform and its distribution are shown in Fig. 8, which justifies the AWGN model used in section A. The histogram is found to be a Gaussian distribution with a standard deviation of 8 mV.

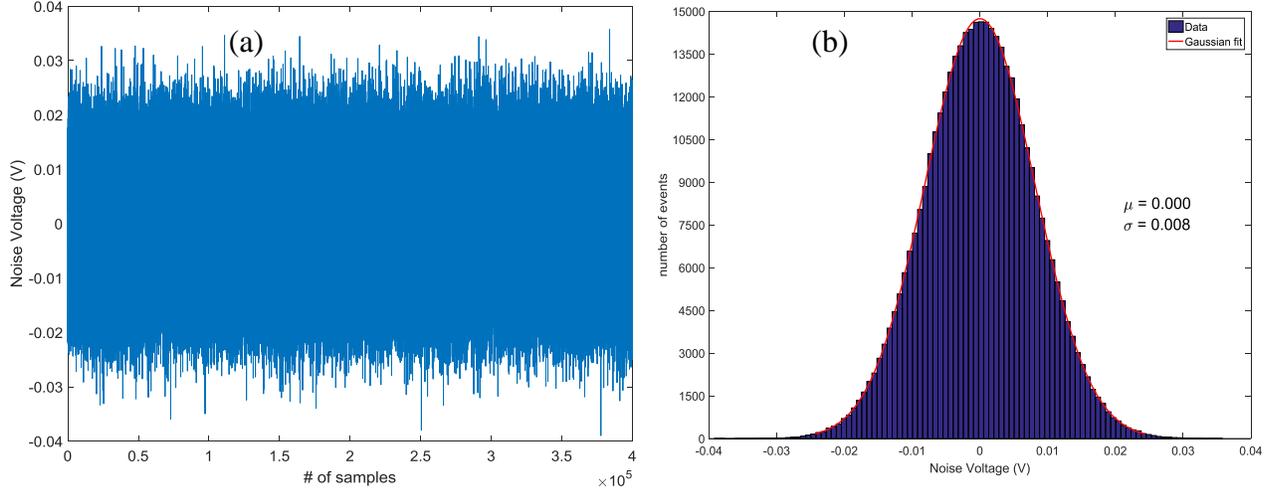

**FIG. 8. Waveform (a) and histogram (b) of noise with 400000 samples. The standard deviation σ is 8 mV.**

Based on this method, it is found that all measured channels (cavity 2, 3, 5, 6) exhibit a similar level of noise (~8 mV RMS) and the variation from channel to channel is at the sub-millivolt level. The SNR can be estimated from the separated signal and noise. With an accelerating gradient of 22 MV/m, the SNR for HOM 1 at cavity 2 of module A1 is approximately 22 dB in contrast to being less than 10 dB for channel 2 during the experiments. The 10 dB difference is mainly due to the different HOM power from the two HOM couplers. In order to compare the simulation and the measurements more accurately, the signals from the simulation are scaled according to the power levels in the measurements.

C.    **Experimental and Simulation Results**

C.1 Comparison between HOM measurements and simulation

Based on the experimental setup described in section B.1, measurements were made at the cavity 2 of injector module A1 of the E-XFEL. The phase of the RF field was varied from 0°, -5° and 5°, with a beam charge of 0.5 nC and the accelerating gradient of ~22 MV/m. For each phase, 25 measurements are made. The resolution obtained is 0.12° and the result is shown in Fig. 9.

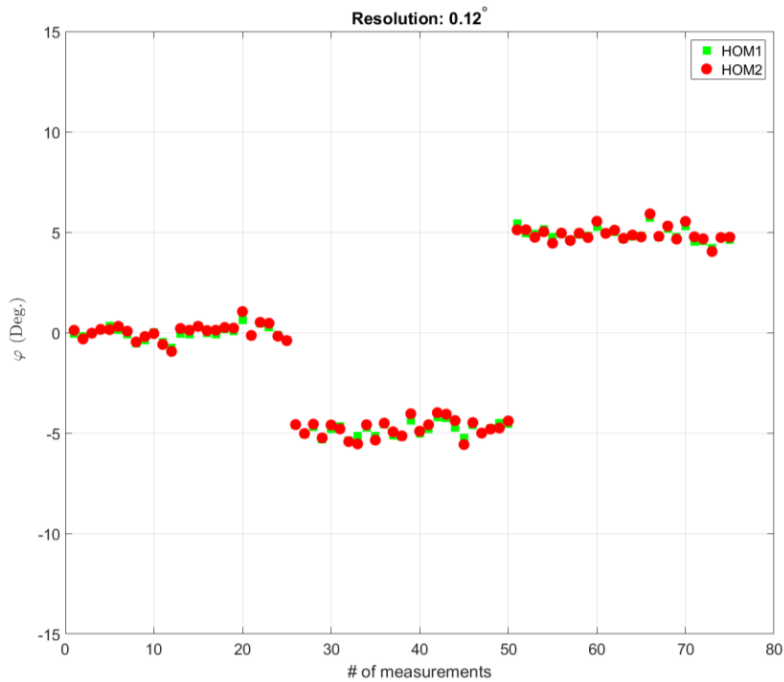

**FIG. 9. The phase $\varphi$ calculated for modes 8 and 9 is shown versus the number of measurements made. For each RF phase value, 25 measurements are made. The beam charge is 0.5 nC and the accelerating gradient is ~22 MV/m. The resolution is estimated to be 0.12°.**

The beam charge was varied between 0.1 nC and 1 nC. The simulation data was scaled according to the measured signal strength based on the method described in section B.2. The phase of the 1.3 GHz signal is changed by -5°, 0° and 5° with 25 calculations per phase value. The comparison between the measurement and the simulation results is shown in Fig. 10.

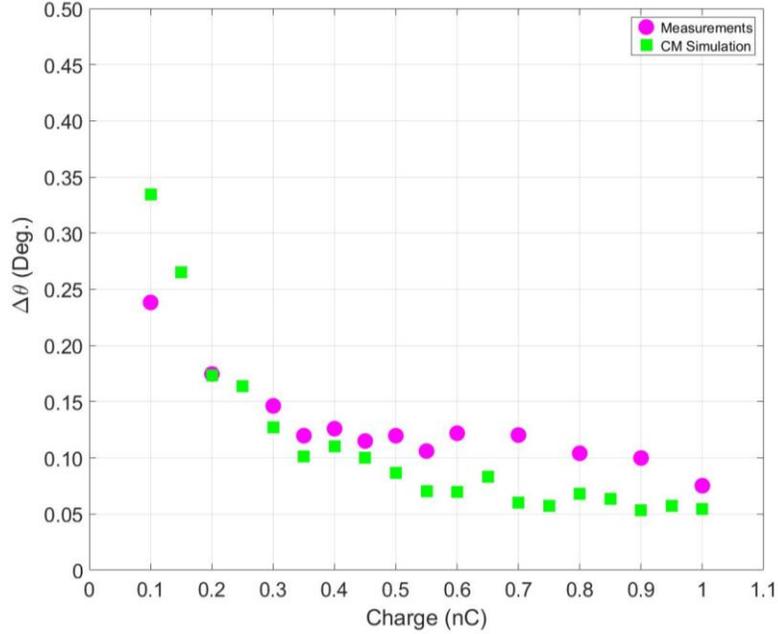

**FIG. 10. Resolution comparison between results from measurements and circuit model simulations (CM).**

The difference is generally below 0.05° except at 0.1 nC where the discrepancy is close to 0.1°. This is due to the fact that the resolution is more sensitive to the charge fluctuations at low charge.

### C.2 Comparison between HOM and LLRF measurements

For the LLRF system, the beam phase is inferred from the 1.3 GHz RF field inside the cavity by means of a field probe. This phase is referred to as probe phase later. The measured phases are readily available from DOOCS. The HOM-based phase, which is based on the experimental setup described in Fig. 5, is called HOM phase.

For comparison, we changed the beam phase from -10° to 10° with a step of 1°. The correlation between HOM Phase 1 and 2 is shown in Fig. 11. The linear dependence indicates the good agreement of the same beam phase measurement from two independent channels.

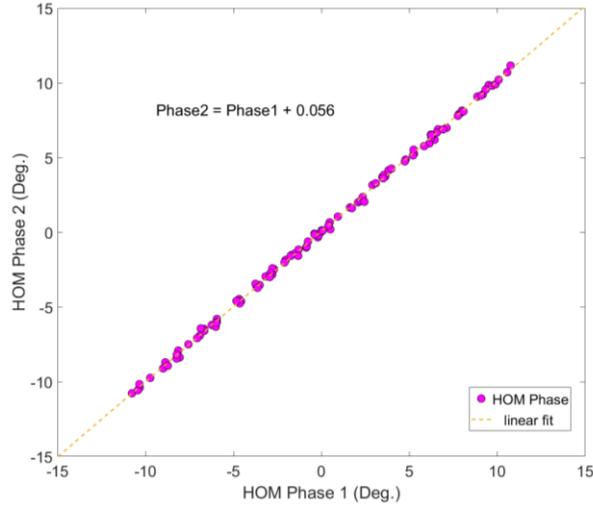

**FIG. 11. HOM phase response to the RF phase change. A linear dependence of the two HOM phases indicates a good agreement (with the coefficient of determination above 0.99).**

During the measurements, the vector sum phase (VS phase) of 32 cavities [16] and the probe phase of a cavity were also recorded. A comparison between the VS phase and HOM phase is shown in Fig. 12 (a). The RMS error between the two is approximately 0.4°. The comparison between the probe phase and HOM phase is shown in Fig. 12 (b). The RMS error is approximately 0.3°.

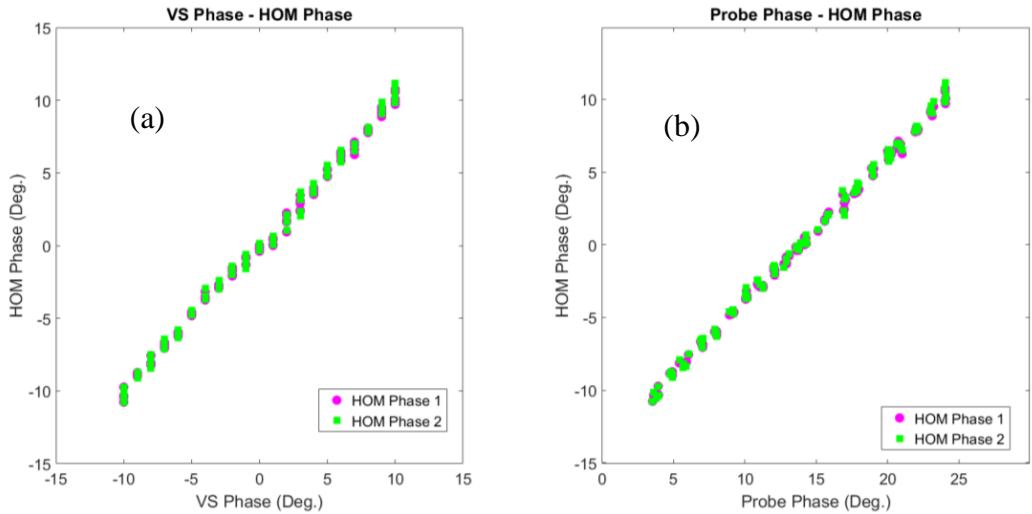

**FIG. 12. HOM phase versus the vector sum phase (a) and versus the probe phase (b).**

The RF phase is fixed at 0° and 50 measurements were made. The histogram of the HOM phase from channels 1 and 2 and the probe phase are shown in Fig. 13.

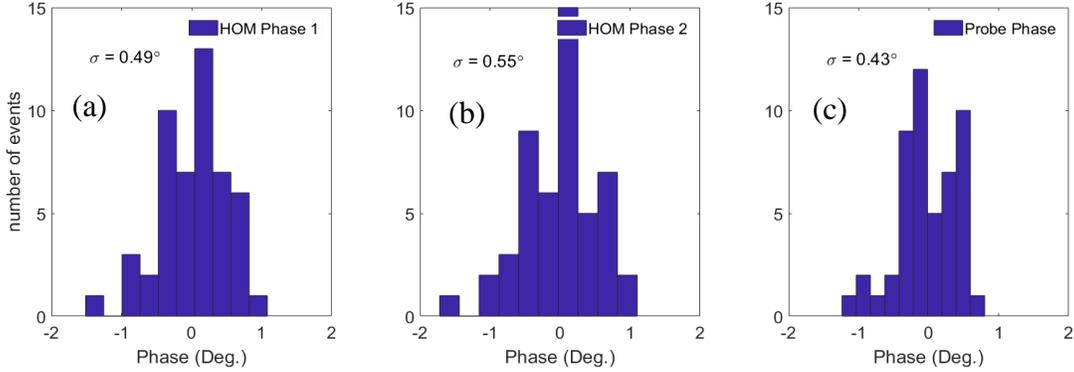

**FIG. 13. Histogram of HOM and probe phases when the RF phase is set to zero.**

The standard deviation of HOM 1 and HOM 2 are 0.49° and 0.55° respectively. As a comparison, the standard deviation from the probe phase is 0.43°. The HOM phase is consistent with the probe phase statistically. The result is expected to be improved when the measurement system is fully synchronized in the future.

### C.3 Theoretical resolution limit

The resolution of the HOM phase measurements has a strong dependence on the noise level present in the system. The theoretical resolution limit is estimated based on the assumption that only thermal noise is present. The smallest measurable thermal energy, $U_{th}$, is [17],

$$U_{th} = \frac{1}{2} k_b T, \qquad (4)$$

where $k_b$ is the Boltzmann constant, and $T$ the temperature, assumed to be 300 K. The amount of energy is thus approximately $2.07 \times 10^{-21}$ J or 0.013 eV.

The energy deposited into mode 8 is:

$$U_8 = kq^2 = 0.6 \frac{V}{pC} \times (500 \; pC)^2 = 0.94 \; TeV, \qquad (5)$$

where $k$ is the point charge loss factor of the mode and $q$ is the charge. A fraction, $\beta$, of the deposited energy can be coupled out through the HOM coupler [9], which is assumed to be 0.5. The SNR based on mode 8 and the thermal detectable noise power is 136 dB. The resolution at this SNR, based on equation 2 at 20 GS/s, is $6 \times 10^{-8}$ degree. The best resolution we have obtained so far experimentally is 0.03°. The SNR is approximately 22 dB. The discrepancy between the obtained resolution and the theoretical limit is due partially to the RF components used during the measurements. The SNR can be improved by amplifying the signal and limiting the noise with RF amplifiers and filters. This will entail a significant effort in the design of the HOM-based diagnostics electronics. However the required SNR (35 dB) is achievable with current technology. The electronics for this purpose are under development in the lab.

### IV. Summary

For a cavity with cylindrical symmetry, the beam excited monopole modes show no dependence on the beam offset and are only dependent on its charge. This characteristic makes them attractive for beam phase monitoring. Both the accelerating mode and the HOMs can be measured together from the HOM coupler via the same RF cable, which makes an on-line direct beam phase monitoring possible.

Based on a coupled circuit model, which has been implemented in the Simulink software, we simulated the excitation of HOM signals and from there the beam phase could be extracted. The resolution of the beam phase determination depends exponentially on the SNR. To fulfil the beam phase accuracy of 0.01°, at least 35 dB signal to noise ratio is required based on the simulation results. However, at the first stage we plan to use this principle to monitor the long term RF phase drift and also to help decouple the phase jitter sources in the E-XFEL injector. These applications require a phase accuracy of no more than 0.1° which is well within that of the current experimental conditions.

Based on a broadband setup, we measured the beam phase at the injector module of the E-XFEL. The resolution routinely obtained is ~0.1° with 0.5 nC beam charge and 20MV/m accelerating gradient. It should be mentioned that the best resolution observed experimentally was 0.03°.

The noise level of the experimental data is estimated based on the SVD method. A fraction of noise was added to the simulation data based on this estimation. With the superimposition of noise, the results from experiment and simulation are consistent with each other and the discrepancy is generally below 0.05°. The measurements are also compared with the readouts from the LLRF system with a good agreement.

The monitor presented in this paper is the first type that can directly and online monitor the phase between the electron beam and RF field inside a cavity. This will provide valuable diagnostic information for the LLRF system, such as the long-term monitoring of stability of the beam phase. The electronics for the European XFEL are under development.

**ACKNOWLEDGMENTS**

We are pleased to acknowledge the support of the FLASH and E-XFEL crew members during the experiments. Special acknowledgement is given to Christian Schmidt and Holger Schlarb for discussions on the beam phase measurements and the LLRF system.
The work is part of EuCARD$^2$, was partly funded by the European Commission, GA 312453.

**APPENDIX A: Procedure of HOM-based phase determination**

This section describes the procedure used to determine the beam phase w.r.t. RF from the HOM couplers:

1. Decomposition of the signal

    The signal $x(t)$ from a HOM coupler contains the 1.3 GHz accelerating mode from RF source and the beam-excited HOMs. Here we only consider the modes in the 2$^{nd}$ monopole band. The signal can be decomposed into sinusoidal terms $x_{si}(t)$ and $x_{ci}(t)$ according to

$$x_{si}(t) = x(t)\sin(\omega_i t), \qquad \text{A.1}$$

$$x_{ci}(t) = x(t)\cos(\omega_i t), \qquad \text{A.2}$$

where $\omega_i$ is the angular frequency of each mode in the signal and has to be determined beforehand. The amplitude and phase of each mode $i$ is carried by $x_{si}(t)$ and $x_{ci}(t)$. An example is shown in Fig. A.1 for $x(t)$, $x_{si}(t)$ and $x_{ci}(t)$.

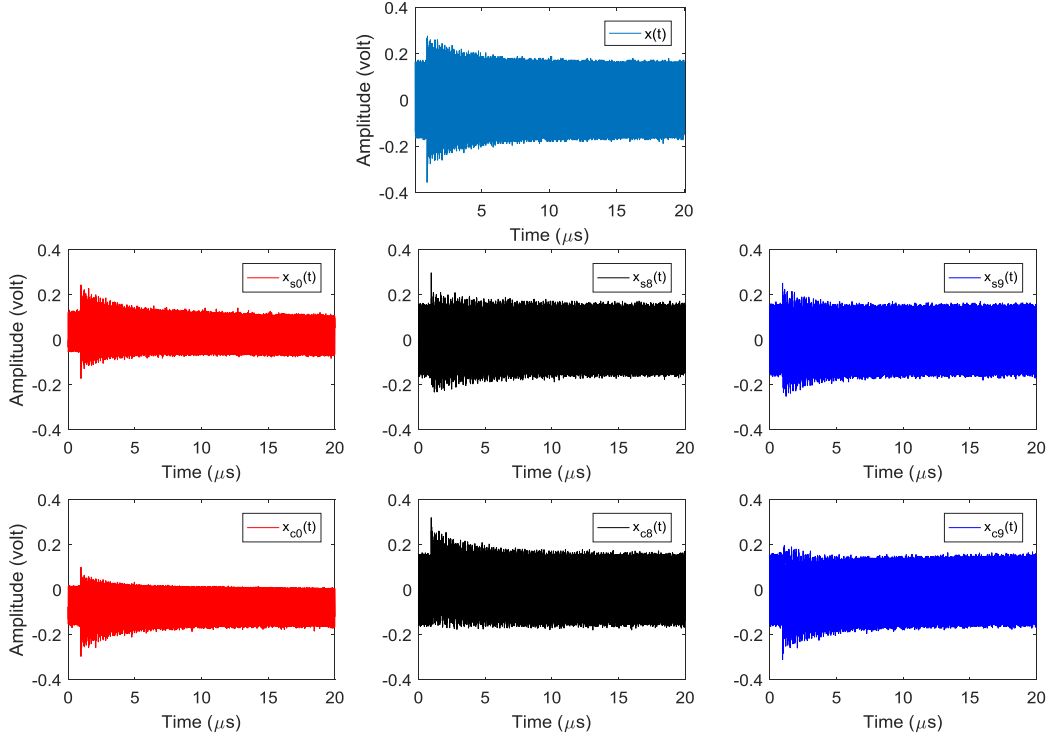

**FIG. A.1. Measured signal *x(t)*, and its components $x_{s0}(t)$ and $x_{c0}(t)$ etc.. $i = 0$ denotes the fundamental 1.3 GHz mode, $i = 8$ and 9 modes 8 and 9 in the second monopole band.**

After the operations in equations A.1 and A.2, $x(t)$ is decomposed into DC and higher frequency components. This is essentially a digital down converting process.

2. Determination of phase for each mode

   The phase of each mode inside the signal $x(t)$ is determined by using equation A.3. By integrating the signal, the DC part accumulates while the higher frequency part is filtered out:

   $$\varphi_i = \tan^{-1}\left(\frac{\int_0^T x_{ci}(t)dt}{\int_0^T x_{si}(t)dt}\right). \tag{A.3}$$

   Taking into account the phase delays from the cavity to the measurement device, the calibration phase $\varphi_{ical}$ is subtracted from $\varphi_i$:

   $$\delta\varphi_i = \varphi_i - \varphi_{ical}. \tag{A.4}$$

3. Determination of beam phase relative to the 1.3 GHz signal

   After calibration, the phase of each HOM is converted into a time delay based on the different frequency of each mode,

$$t_i = \frac{\delta\varphi_i}{\omega_i}. \quad \text{A.5}$$

The average arrival time is defined as,

$$t_a = \sum_{i=2}^{N} w_i t_i, \quad \text{A.6}$$

where $N$ is the number of HOMs used in the calculation and $w_i$ is the normalized power of mode $i$. The phase of the accelerating mode signal at 1.3 GHz relative to $t_a$ can be calculated with

$$\varphi_{acc} = \omega_{acc}(t_1 - t_a), \quad \text{A.7}$$

where $\omega_{acc} = 2\pi \times 1.3 \times 10^9$ rad·s$^{-1}$ is the angular frequency of the accelerating mode.

Based on steps 1-3, the beam phase relative to RF field can be calculated. The algorithm is implemented in a MATLAB script.